# Dynamic Scheduling of Virtual Machines Running HPC Workloads in Scientific Grids

Omer Khalid
CERN
Geneva, Switzerland
Omer.Khalid@cern.ch

Ivo Maljevic
Soma Networks
Toronto, Canada
IMaljevic@somanetworks.com

Richard Anthony
University of Greenwich
London, United Kingdom
R.J.Anthony@gre.ac.uk

Miltos Petridis
University of Greenwich
London, United Kingdom
M.Petridis@gre.ac.uk

Kevin Parrott
University of Greenwich
London, United Kingdom
A.K.Parrott@gre.ac.uk

Markus Schulz
CERN
Geneva, Switzerland
Markus.Schulz@cern.ch

## ABSTRACT
The primary motivation for uptake of virtualization has been resource isolation, capacity management and resource customization allowing resource providers to consolidate their resources in virtual machines. Various approaches have been taken to integrate virtualization in to scientific Grids especially in the arena of High Performance Computing (HPC) to run grid jobs in virtual machines, thus enabling better provisioning of the underlying resources and customization of the execution environment on runtime. Despite the gains, virtualization layer also incur a performance penalty and its not very well understood that how such an overhead will impact the performance of systems where jobs are scheduled with tight deadlines. Since this overhead varies the types of workload whether they are memory intensive, CPU intensive or network I/O bound, and could lead to unpredictable deadline estimation for the running jobs in the system. In our study, we have attempted to tackle this problem by developing an intelligent scheduling technique for virtual machines which monitors the workload types and deadlines, and calculate the system over head in real time to maximize number of jobs finishing within their agreed deadlines.

## Categories and Subject Descriptors
D.2.7 [**Software Engineering**]: Distribution, Maintenance, and Enhancement – *portability*; Metrics – *performance measures*;

## 1. INTRODUCTION
The biggest challenge for running HPC jobs in the virtual machines (VM) on the Grid [2] lies in how significant the virtualization overhead is since the virtualization technology became an established desktop tool [15], and whether jobs with tight deadlines could meet their obligation if resource providers were to fully virtualizes their worker nodes.

Given this potential, we decided to investigate how this technology could benefit ATLAS [12] (one of European Center of Nuclear Research - CERN's high energy physics experiments) grid infrastructure and improve its efficiency by simulating its HPC jobs on virtual machines for tight deadlines of completion.

This poses a particular challenge in scientific grids such as LCG[1] that have to serve the needs of diverse communities often with competing and opposite demands. Once virtualization is enabled, the next step is to minimize the virtualization overhead incurred by the jobs, as they have to run longer to complete due to extension in their duration. This is simpler to manage in commercial clouds such as Amazon's Elastic Cloud Computing (EC2)[2] [23] or scientific clouds like Nimbus[3] [16] where user have clear understanding that they would be paying for per hour usage and their SLA would terminate when they stop to pay.

Since most of the scientific clouds cater the needs of different HPC communities, and have strict policies that it would kill the user jobs when they over run their allocated time limits, this can result in significant reduction in utilization efficiency. Our study attempts to provide a way forward to address the above mentioned challenges in a way which is transparent to the users with out letting them know that their jobs are run on the virtual machine and tries to optimize the job success rate in a virtual machine.

---

[1] Large Hadron Collider Computer Grid (LHC) and Open Science Grid (OSG).

[2] Amazon provides an on demand cloud computing service where user pays for per hour usage of virtual machines.

[3] Nimbus is a cloud computing service provided by University of Chicago to academic organizations.

## 2. MOTIVATION AND BACKGROUND

The problem we are seeing is that more focus and attention have been given to *utility computing* and *cloud computing* but leaving out the very important question of how to schedule mixed workloads with competing requirements at the machine level. There is lot of effort being made in standardizing and hiding the complexity of resource management and allocation at a cluster level and exposing it to the end users as *cloud*.

To serve to diverse user communities with often competing Quality of Service (*QoS*) requirements for their jobs/virtual machines, some jobs being more CPU or memory intensive than the others and vice versa, requires a dynamic and intelligent resource scheduling which is adaptive as the nature of workloads at any given moment changes. *QoS* varies from different utility context such as its different for EC2 user community as compared to the users of particle physics community.

### 2.1 Virtualization

The virtual machine technology has a long history. IBM first successfully implemented it in its VM/370 operating system that allowed user to time-share hardware resources in a secure and isolated manner. With the rise of powerful desktop computing, the present day virtualization technology [6, 7] provides the following benefits:

- o Flexibility and Customization: Virtual machines could be configured and customized with specific software such as applications, libraries etc for different LHC experiments without directly influencing the physical resources. This decouples the environment from the hardware, and allows fine-grain customization to enable support for jobs with special requirements such as root access or legacy applications.

- o Security and Isolation: Virtualization adds an additional layer of security as activity taking place within one VM is independent and isolated from the other VM's by first preventing a user of one VM affecting the performance or integrity of other VM's, and secondly limiting the activity of a malicious user, if a VM is compromised, to be restricted to that particular VM. This allows the underlying physical resource staying independent and secure in event of a security breach, and the compromised VM could be shutdown without affecting the whole system.

- o Migration: VM's are only coupled to the underlying hypervisor, due to difference of image formats, but stays independent from the physical machine. This capability is particularly useful if an executing job have to be suspended and migrated to another physical machine or site. This capability allows migrating virtual machine image with the saved state for a job and poses very few constraints on the site.

- o Resource Control: Virtualization allows fine-grained control to the resource providers to allocate well-defined and metered quantities of physical resources (CPU, network bandwidth, memory, disk) among multiple virtual machines. This leads to better utilization of server resources and could be dynamically managed to match demand-supply profile among competing virtual machines. It also enables fine-grained accounting of resource consumption by the virtual machines, and thus fits very well with the Virtual Organization (VO) resource control policies.

### 2.2 Architecture

Since ATLAS experiment uses PanDA [13] software framework to submit jobs to the grid. In our previous experiments, we demonstrated how such a existing Grid application framework is modified to deploy grid jobs in virtual machines [8, 9, 10], This is illustrated in the following figure 1.

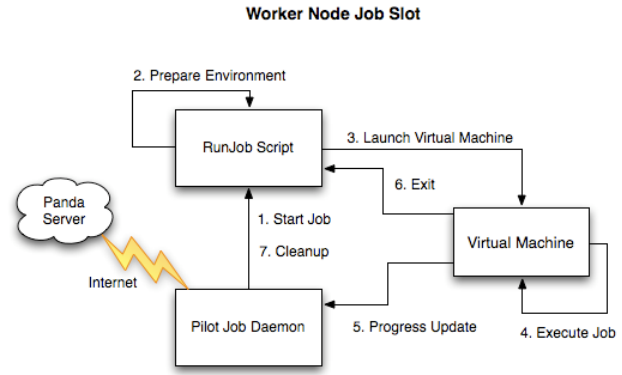

**Fig 1. Once the pilot job has started, it launches the *runjob* script, which requests the virtual machine container from the *deployment agent* and starts the job execution. Once started, the virtual machine updates the main pilot job of its status and upon job failing/termination; the *runjob* script requests the shut down of the virtual machine**

### 2.3 Simulation Model

Once the virtualization was enabled for the PanDA pilot job framework [14], we developed a simulator to deal with the constraint placed on our system as we couldn't run the all the ATLAS jobs in virtual machine which could run up to 24hrs each. Since we needed to run thousands of jobs to test the algorithm, thus simulation was only realistic way to do it.

In our simulation model, the algorithm intelligently schedulers the jobs and learn over time about the missed deadlines under various conditions and try to predict whether the job would be meeting its deadline and if not then take appropriate measures to improve its success chances.

Further more, since deadlines miss rate is an emerging property of the system depending on the uncertain behavior of concurrent jobs and profile of the jobs in the global job queue, which alters the virtualization overhead of the system dynamically.

We introduced two transient variables in the system to allow the scheduling algorithm to respond to the system properties.

$x$ factor is a ratio of a job $i$ that is projected to miss its current deadline, and is determined by:

$$x_i = \frac{(job\ duration\ remaining - time\ to\ deadline)}{job\ duration\ remaining} \quad (1)$$

The first, and easiest method, is to set an acceptance threshold such that when $x_i < x$ threshold $(X)$ jobs are accepted and rejected otherwise. The basic idea behind this approach is that it is expected that that acceptance of jobs beyond a certain threshold would be counter-productive as most of them would fail.

The algorithm can be easily described as following:

```
if x_i < X:
        accept_job()
else:
        reject_job()
```

The drawback of the threshold method is that once set, the threshold does not change, and if the system behavior changes over time, initially selected threshold is no longer valid. To avoid this problem, we use another approach, where threshold dynamically adapts to the currently existing conditions while trying to keep the failure rate close to a selected target failure rate. The threshold value for job acceptance is lowered if the failure rate increases and vice versa. The update step has been quantized to a small value $\Delta x$ in order to avoid fast threshold changes if the difference between the measured and targeted failure rate is large. The optimal value for the step $\Delta x$ has been determined through experimentation.

It has been observed that the control loop used in this algorithm can become unstable, so as a safeguard, threshold values can be varied only within a certain range $X_{min} < X_{Thresh} < X_{max}$.

$$X_{Thresh_i} = X_{Thresh_{i-1}} + \Delta x * | Failure_{target} - F_{measured} | \qquad (2)$$

A third approach we have taken is to calculate cumulative distribution function (CDF) of the success rate and use it to try to select the threshold value dynamically in a such way that $X_{Thresh}$ always corresponds to the probability that is expressed in percentage of the jobs are successfully completed.

It have to be noted that an arriving at an optimal value of $X_{Thresh}$ is important since keeping it too high will lead to more jobs being accepted but eventually missing their deadlines or keeping it too low will kill the jobs pre-maturely. Both scenarios will results in lower system utilization and performance.

## 3. CASE STUDY

Our test bed consist of 2 servers each with SMP dual-core Intel CPU's running at 3 GHz, 8 GB RAM and 1 Gbit/s NIC's running Xen 3.1 [1, 3] and Scientific Linux CERN 4 (SLC4). Since the present study is based on simulation results, we used the same machine resources configuration for the virtual server as of the physical servers with 4 CPU and 8GB RAM. For the training phase, to derive some core parameters we ran the simulator for job queue length of 10,000 hours of workload while for the steady phase the job queue length was 100,000 hours of workloads.

Since there are many different input parameters in the system such as resource ratio (memory to CPU), frequency of the scheduler, deadline buffer that was set to 5% of the job duration, alpha and delta values for the adaptive x algorithm. We first ran the simulator in the training mode to establish optimum values for the above-mentioned parameters before running the actual simulation.

Once the key optimization parameters were measured through the training phase, we ran the simulation for 100k hours of workloads (mixed jobs with low, high CPU and memory requirements) for different set of configuration (alg_1, alg_2, alg_3, alg_4, alg_5) to measure that how HPC workloads will perform when ran on virtual machines.

The algorithm was tested for the following configurations:

1. **Physical** baseline (alg_1) representing the workload being executed on physical machines as compared to virtual machines.
2. **Virtual Static** (alg_2) represents the configuration where no intelligent virtualization overhead (VO) management was done
3. **Virtual Dynamic** (alg_3) represents the workload executed in virtual machine with dynamic VO management
4. **Virtual Dynamic Adaptive** (alg_4) configuration was run with dynamic VO using adaptive algorithm for $x$ threshold.
5. **Virtual Dynamic Statistical** (alg_5) configuration was run with Dynamic VO but using CDF to adapt the $x$ threshold for the executing work loads.

Figure 2 shows the performance results for the above-mentioned algorithms. For each algorithm, there are two lines (labeled); one for over all job success rate (solid line) and the second for deadline miss rate (broken line).

It was observed that alg_1 had the best performance in overall system success rate with the minimum job deadline miss rate and since one core objective of this study have to be develop an virtual machine optimization algorithm that could come close to this performance. With out such an optimization, static VO (alg_2) lead to worst performance of 0.42 with deadline miss rate of 0.58. Where as, alg_3 showed a considerable performance jump from 0.42 to 0.78 by 85% for success rate but when our adaptive algorithm activates (for both alg_4 and alg_5 configuration), it further improved the job success rate from 0.78 to 0.84 by 7.7%.

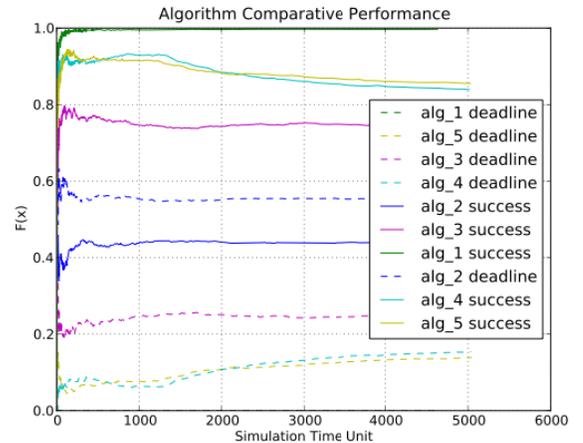

**Fig 2. Comparative performance of different algorithms tested for the simulation**

Similar gains were observed for meeting job deadlines on time while running in virtual machines where alg_4 and alg_5 had a deadline miss rate of ±0.17 which is 26% less than alg_3's 0.23 deadline miss rate.

For alg_4, as show in figure 3, after learning phase (1k simulation time unit), as the failure rate rises, $X$ is raised to the ceiling of 0.9 very quickly and it drops to 0.1 while the failure rate keeps on floating. Where as, for alg_5, as show in figure 4, $X$ rarely touches the ceiling of 0.9 even though the fluctuations in the failure rate are similar to alg_4 failure rate. It consistently converges to *x threshold* values between [0.1, 0.3]. Further more, alg_5's failure rate never went above 0.79 where as alg_4's peaked at 0.85.

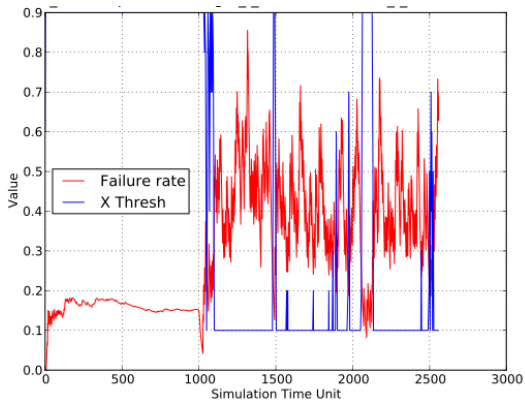

**Fig 3. X threshold and failure rate evolution for adaptive *x* algorithm used in alg_4 configuration**

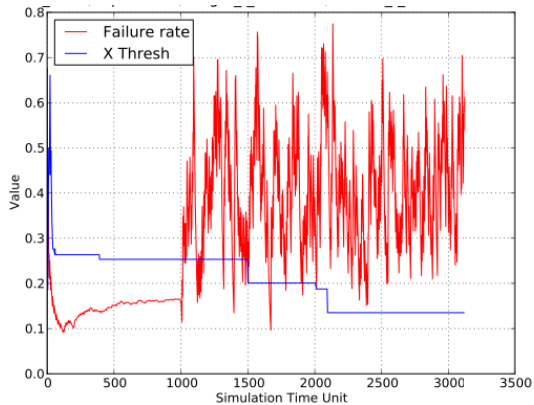

**Fig 4. X threshold and failure rate evolution for statistical driven algorithm used in alg_5 configuration**

## 4. RELATED WORK

Xen had historically 3 schedulers built in and the work did by Cherkasova et al is particularly interesting but they mainly focused on interactive workloads, and compared different Xen schedulers [4, 5].

Our present work is an extension to approach that has been built upon the existing experience and know-how of the infrastructure such as the LHC[1] worldwide grid for the physics and scientific community, so that existing deployment frameworks could be used to enable on-demand virtual machine based job execution on the Grid.

Our work is complement to scheduling approaches taken by VSched [11] as their focus have been on virtual machines running real-time and user-interactive (UI) applications which requires scheduling to keep the UI latency as low as possible while our workloads are HPC and ran in batch mode with no user interactivity. Similar work have been done to re-shape jobs and to resize virtual machines dynamically according to the load on the system but this work have been primarily target towards parallel job execution [21]. Lingrand et al [22] have taken alternative approach to apply optimization technique to job submissions and it would be very interesting to explore this further in the context of job submission to virtual machines.

This work is particularly interesting in the scope of *hotspot* management in the datacenters where a lot of efforts have been invested to dynamically move the virtual workloads around in the datacenter to manage temperature of the servers [18, 20], and thus to reduce data center operational cost and consolidating workloads on fewer servers under peak loads while standing by the unutilized server. In such circumstances, retaining the network properties of the migrated VM and restarting the job executing in the VM from where it was stopped addressed in the following studies [17, 19]. Since our deadline-aware scheduler dynamically monitoring the job and could live-migrate it to another server if it could meet the job's deadline. This is a very interesting research for further investigation but it remains outside the scope of our study.

## 5. CONCLUSION

In this paper, we have described a dynamic and adaptive real-time virtual machine scheduling technique for HPC workloads when ran on the Grid and to optimize the performance impact of virtualization on their deadline obligations. We examined the impact of various techniques to calculate virtualization overhead and their impact on job performance by using $X$ threshold and measuring failure rate in real time to take dynamic scheduling decisions of whether letting the job continue based on its chances to succeed or to terminate at that point in time. As shown in the above results, this increases overall job throughput in the system with higher resource utilization rates by pre-maturely terminating the executing jobs, which would have never completed otherwise.

Our scheduler can also be integrated with a high-level batch/cluster scheduler where each node have different resource composition, then it would be the responsibility of the cluster scheduler to pick the best resource combination for any given job while node level VM scheduler optimizes jobs success rate.

---

[1] Large Hadron Collider (LHC) is at CERN, Switzerland

## 6. ACKNOWLEDGMENTS
We are grateful to ATLAS Computing and PanDA experiment and collaboration in providing support to us during our research and answering our long emails and letting us having access to PanDa experiments historical data. We also thank to Angelos Molfetas for his detailed feedback on the paper, and to Kate Keahey for brainstorming to evaluate some of the key concepts presented in this paper.